\documentclass[aps,prd,twocolumn,showpacs,showkeys,preprintnumbers,amsmath,amssymb]{revtex4}
\usepackage{amsmath}
\usepackage{amssymb}

\begin{document}

\title{Galilean Covariance and The Gravitational Field }
\author{S. C. Ulhoa }
\email{sergio@fis.unb.br}
\affiliation{Instituto de F\'isica, Universidade de Bras\'ilia
70910-900, Brasilia, DF, Brazil}

\author{Faqir C. Khanna}
\email{khanna@phys.ualberta.ca }
\affiliation{Physics Department, Theoretical Physics Institute,
University of Alberta Edmonton, Alberta T6G 2J1 Canada
TRIUMF 4004, Westbrook mall, Vancouver, British Columbia V6T 2A3,
Canada}

\author{A. E. Santana}
\email{asantana@fis.unb.br}
\affiliation{Instituto de F\'isica, Universidade de Bras\'ilia
70910-900, Brasilia, DF, Brazil}
\date{\today}

\begin{abstract}
The paper is concerned with the development of a gravitational
field theory having locally a covariant version of the Galilei
group. We show that this Galilean gravity  can be used to study
the advance of perihelion of a planet, following in parallel with
the result of the (relativistic) theory of general relativity in
the post-Newtonian approximation.
\end{abstract}


\maketitle
\section{Introduction}
\noindent

Since the birth of general relativity, several studies have been
addressing the problem of an analogous (n-form) formulation for
the non-relativistic theory of
gravitation~\cite{Havas,kunzle1,kunzle,Duval,Carter1,Carter2,Carter3},
which has as a fundamental structure the Galilei group. Beyond
that, the interest in such a covariant description of the Galilei
physics lies  in the fact that some phenomena are restricted to
Galilean regime. A known example is the superfluidity phenomenon,
existing at low velocity, only~\cite{Landau1}. Particularly in
cosmology, in order to understand the large scale structures of
the universe, the Newtonian gravity is required. Besides that, the
rotation curve of galaxies is obtained with a Newtonian
formalism~\cite{Brihaye}, which was the first step to point to the
existence of dark matter~\cite{Persic,Gallagher,Faber}. In a
general sense, the Newtonian gravity theory is a natural choice to
verify some insights on gravitation; thus if new ideas emerge from
the development of general relativity, or even from a better
theory than that, it has to be tested in order to reproduce the
known results. In this sense, a geometric formulation of gravity
based on the Galilei group (a Galilei gravity theory) may be of
interest; and this is one of our goal here.

For such a purpose, it is important to develop a covariant form of
Galilean transformations, since Galilei group acts as the symmetry
group of Newtonian theory~\cite{kunzle1,Duval}. This approach has
been achieved by considering the space-time transformation in the
light-cone of a 5-dimensional Minkowski
space-time~\cite{Duval,taka1,taka2,taka3,taka4}. The 5-momentum
vector is interpreted physically by considering 3 components
describing the Euclidian momentum; one component standing for
energy and the fifth component describing mass. In terms of space
canonical coordinates, one has three components for the space
coordinates; one coordinate  for time and the fifth component is
associated to velocity. The consequence has been several
developments for the non-relativistic classical and quantum field
theory~\cite{taka1,taka2,taka3,taka4,taka5,taka6,Montigny,Santos,Kobayashi}.

In this context of non-relativistic covariant physics, Duval et
al~\cite{kunzle} have addressed the problem of the gravitational
field, using Bargmann structures, rather than Galilei group. In
another direction, Carter and Chamel~\cite{Carter1} adapted the
procedures used in general relativity for application in purely
Newtonian framework in order to provide new insights other than
that of 3+1 decomposition of space-time. The formalism have been
applied in the construction of a Newtonian fluid model to treat
effects of superfluidity in neutrons stars~\cite{Carter2,Carter3}.
Here we follow a different perspective, avoiding the use of such a
decomposition, since we work in the five dimensional space
exploring the Galilei group. We develop a geometric description of
a Galilean gravity parallelizing the usual general relativity. As
an application we study the advance of perihelion of a planet and
compare it to the results derived in the post-Newtonian version of
the theory of general relativity.

Although our approach is quite close to the general relativity,
they differ from each other by the fact that one is locally
Lorentzian and the other is Galilean. The structure is the same
and the equations will assume a similar tensor form, but the
physical meaning is different. Since Galilei transformations are
described in five dimensions as linear, similar to the Lorentz
group, the tensor formalism developed in the context of general
relativity, may be used. As a central result, our conclusions are
genuine manifestations of the Galilean gravity and do not follow
approximations of any kind of Einstein's equations, as it is the
case, for example, in the post-Newtonian
approximation~\cite{Weinberg}, which is an expansion in terms of
$1/c$, where $c$ is the speed of light.

The paper is organized in the following way. In section 2, we set
forth the notation and discuss briefly some aspects of the
covariant formulation of Galilean transformations. In section 3 we
establish a geometrical formulation of Galilean gravity and in
section 4, we apply it to the case of Schwarzschild-like line
element to calculate the precession of the perihelion of a planet.
In section 5, we present some concluding remarks.

\section{Covariant Galilean Transformations}
The Galilean transformations are given by
\begin{eqnarray}
\textbf{x}'&=& R\textbf{x}-\textbf{V}t+\textbf{a} \nonumber\\
t'&=&t+b\,, \label{0}
\end{eqnarray}
where $R$ is a 3-dimensional Euclidian rotation, $\textbf{V}$ is
the relative velocity defining the Galilei boost, $\textbf{a}$
stands for a space translations and $b$ a time translation. In
this realm of Galilean symmetries describing low velocities
process, one can introduces a linear space-time tensor structure
by  noticing the following.

In non-relativistic physics, the dispersion relation for a free
non-relativistic particle is given by $E=\mathbf{p}^{2}/2m,$ where
$E$ is the energy, $\mathbf{p}$ is the 3-dimensional momentum and
$m$ is the mass.  This dispersion relation can also be written as
\begin{equation}
\mathbf{p}^{2}-2mH=0.  \label{disp5}
\end{equation}
Now let us consider the physical observable describing  momentum
as a quantity consisting of five entries, that is $p^{\mu
}=(\mathbf{p},p^{4},p^{5}),$ where $\mu =1,...,5, $ $\mathbf{p}$
is standing for the 3-vector momentum, $p^{4}=E/c'$ is the energy,
and $p^{5}=c'm$ is mass. Here $c'$ is a constant with units of
velocity. We take $c'=1$. Using this notation, and in order to
recover Eq.~(\ref{disp5}), we write a general 5-dimensional
dispersion relation, i.e. $p_{\mu }p^{\mu }=p^{\mu }p^{\nu
}{\eta}_{\mu \nu }=\mathbf{p}^{2}-2p^{4}p^{5}=k^{2},$ where
\begin{equation}
\eta_{\mu\nu}= \left(
  \begin{array}{ccccc}
    1 & 0 & 0 & 0 & 0 \\
    0 & 1 & 0 & 0 & 0 \\
    0 & 0 & 1 & 0 & 0 \\
    0 & 0 & 0 & 0 & -1 \\
    0 & 0 & 0 & -1 & 0 \\
  \end{array}
\right) \,. \label{1}
\end{equation}
This is taken as a metric tensor, that has been introduced in
different ways in the literature, in particular, it was obtained
in a (1+1) theory of gravitation by Cangemi and
Jackiw~\cite{Jackiw}.

Let us define the set of canonical coordinates associated
to $p^{\mu }$, by writing a 5-vector in $\mathcal M$ as $q^{\mu }=(\mathbf{q}%
,q^{4},q^{5}).$ The entries in $q^{\mu }$ are physically
interpreted as follows: $\mathbf{q}$ is the canonical coordinate
attached to $\mathbf{p}$; $q^{4}$ is the canonical coordinate
associated to $E,$ and so it can be considered as the time
coordinate; $q^{5}$ is the canonical coordinate
associated to $m$, and is explicitly given in terms of $\mathbf{q}$ and $%
q^{4}$ according to the corresponding dispersion relation, leading to $%
q_{\mu }q^{\mu }=q^{\mu }q^{\nu }{\eta}_{\mu \nu }=\mathbf{q}%
^{2}-2q^{4}q^{5}=s^{2}.$ Since $p_{\mu }p^{\mu }=0$, we have to
take $s=0$, leading to $q^{5}={\mathbf q}^2/2t$; or
infinitesimally, we obtain $\delta q^{5}={\mathbf v}\cdot \delta
{\mathbf q}/2$. Therefore the fifth component is basically defined
by the velocity.

Let us study in more detail the content of canonical coordinates,
by introducing a symplectic structure in the
 cotangent bundle $T^{\ast }{\mathcal M}$,  through the 2-form $\omega $ ,
\begin{equation}
\omega ={\eta}^{\mu \nu }dq_{\mu }\wedge dp_{v},\quad \,\,\mu
=1,2,\dots 5. \label{c4}
\end{equation}
Defining the vector field,
\[
X_{f}=\frac{\partial f}{\partial p_{\mu }}\frac{\partial }{\partial q^{\mu }}%
-\frac{\partial f}{\partial q^{\mu }}\frac{\partial }{\partial
p_{\mu }},
\]
where $f$ is a $C^{\infty }$ function in the 10-dimensional (phase
space) manifold $\Omega $ with coordinates $(q^{\mu },p^{\nu })$,
we have
\begin{eqnarray*}
\omega (X_{f},X_{h}) &=&\{f,h\} \\
\ &=&dq(X_{f})dp(X_{h})-dp(X_{f})dq(X_{h}), \\
\ &=&{\eta}^{\mu \nu }(\frac{\partial f}{\partial q^{\mu }}\frac{\partial g}{%
\partial p^{\nu }}-\frac{\partial g}{\partial q^{\mu }}\frac{\partial f}{%
\partial p^{\nu }})\ ,
\end{eqnarray*}
where $\{f,g\}$ is the Poisson bracket. Observe that, since
$w(X_{h})=dh$, we have $\{f,h\}=df(X_{h})=\left\langle
df,X_{h}\right\rangle $.

Defining a flow by
\begin{equation}
\partial _{\mu }f=-X_{p_{\mu }}f,  \label{c335}
\end{equation}
where $f(q,p)$ is a real ($C^{\infty })$ density distribution function in $%
\Omega $, then, in terms of components, we have from Eq. (%
\ref{c335}),
\begin{eqnarray}
\partial _{i}f &=&-X_{p_{i}}f\mapsto \partial _{i}f=\{p_{i},f\},  \label{c51}
\\
\partial _{4}f &=&-X_{p_{4}}f\mapsto \partial _{t}f=\{H,f\},  \label{c52} \\
\partial _{5}f &=&-X_{p_{5}}f\mapsto \partial _{5}f=0.  \label{c53}
\end{eqnarray}
Consistency relations are given by Eq. (\ref{c51}) and
(\ref{c53}), while Eq. (\ref{c52}) describes the Liouville
equation. In this way we recover the classical theory in the
Liouville-Poisson representation.

Let us now turn our attention to the set of linear non-homogeneous
transformations in $\mathcal M$ of type $\overline{q%
}^{\mu }=\Lambda_{\ \nu }^{\mu }q^{\nu }+a^{\mu }\ $, leaving
$(dq^{\mu }dq_{\mu }^{\prime })$ invariant. In addition,  we
consider transformations connected to the identity, such
that$|\Lambda|=1$. In this case, for infinitesimal transformations
we have $\Lambda_{\ \nu }^{\mu }=\delta _{\ \nu }^{\mu }+\epsilon
_{\ \nu }^{\mu }$.  Then we identify 15 generators of
transformations. Using the definition
\begin{equation}
\left. \widehat{K}_{\alpha }=i\frac{\partial \overline{q}^{\mu
}}{\partial \alpha }\right| _{\alpha =0}\frac{\partial }{\partial
q^{\mu }},  \label{e12}
\end{equation}
where $K_{\alpha }$ is the generator associated with the group parameter $%
\alpha $ (which also labels the group generators), we have
\begin{eqnarray}
\widehat{J}_{3} &=&-i(\,q^{1}\partial _{2}-\,x^{2}\partial _{1}),
\label{e13} \\
\widehat{J}_{1} &=&-i(\,q^{2}\partial _{3}-\,q^{3}\partial _{2}),
\label{e14} \\
\widehat{J}_{2} &=&-i(\,q^{3}\partial _{1}-\,q^{1}\partial _{3}),
\label{e15} \\
\widehat{G}_{i} &=&\,i(\,q^{4}\partial _{i}+\,q^{i}\partial _{5}),
\label{e16} \\
\widehat{C}_{i} &=&\,i(\,q^{5}\partial _{i}+\,q^{i}\partial _{4}),
\label{e17} \\
\widehat{D} &=&\,\,\,i(\,q^{4}\partial _{4}-\,q^{5}\partial _{5}),
\label{e18} \\
\widehat{P}_{\mu } &=&\,\,\,i\partial _{\mu },  \label{e19}
\end{eqnarray}
where $i=1,2,3$ and $\mu =1,2,...,5$. These generators satisfy the
following commutation relations:
\begin{eqnarray}
\lbrack \widehat{M}_{\mu \nu },\widehat{M}_{\rho \sigma }]
&=&-i[\eta _{\nu \rho }\widehat{M}_{\mu \sigma }-\eta _{\mu \rho
}\widehat{M}_{\nu \sigma
}+\eta _{\mu \sigma }\widehat{M}_{\nu \rho }- \nonumber\\
&-&\eta _{\nu \sigma }\widehat{M}_{\mu \rho }],  \label{e25} \\
\lbrack \widehat{P}_{\mu },\widehat{M}_{\rho \sigma }] &=&-i[\eta
_{\mu \rho
}\widehat{P}_{\sigma }-\eta _{\mu \sigma }\widehat{P}_{\rho }], \\
\lbrack \widehat{P}_{\mu },\widehat{P}_{\nu }] &=&0,  \label{e27}
\end{eqnarray}
where $\widehat{M}_{\alpha \beta }$ ($\alpha ,\beta =1,...,5$) are
defined by

\begin{eqnarray}
\widehat{M}_{ij} &=&-\widehat{M}_{ji}=\varepsilon
_{ijk}\widehat{J}_{k},
\label{e21} \\
\widehat{M}_{5i} &=&-\widehat{M}_{i5}=\widehat{G}_{i},  \label{22} \\
\widehat{M}_{4i} &=&-\widehat{M}_{i4}=\widehat{C}_{i},  \label{e23} \\
\widehat{M}_{54} &=&-\widehat{M}_{45}=\widehat{D}.  \label{e24}
\end{eqnarray}
The commutation relations given in Eqs.~(\ref{e25})--(\ref{e27})
is a Lie algebra, that we denote by $\mathbf{g}$. A  subalgebra of
$\mathbf{g}$ is
\begin{eqnarray}
\lbrack \widehat{L}_{i},\widehat{L}_{j}] &=&i\varepsilon _{ijk}\widehat{L}%
_{k},\,\,\,[\widehat{L}_{i},\widehat{P}_{j}]=i\varepsilon _{ijk}\widehat{P}%
_{k},\,\,\,[\widehat{L}_{i},\widehat{B}_{j}]=i\varepsilon _{ijk}\widehat{B}%
_{k},  \nonumber \\
\lbrack \widehat{B}_{i},\widehat{P}_{4}] &=&i\widehat{P}_{i},\,\,\,[\widehat{%
B}_{i},\widehat{P}_{j}]=i\widehat{P}_{5}\delta _{ij}, \label{alg2}
\end{eqnarray}
corresponding to the Galilei-Lie algebra with the usual central
charge $\widehat{P}_{5}$ describing mass. Notice that here the
central charge arises naturally from the isometry in 5-dimensions.

The dispersion relation $p_\mu p^\mu=0$ defines a Galilean vector
in the light-cone. However, in a more general case we have
\begin{eqnarray}
p_\mu p^\mu&=&p^2-2mE=k^2 \nonumber\\
E+k^2/2m&=&\frac{p^2}{2m}. \label{3.1}
\end{eqnarray}
The constant $k^2$ is absorbed  into the energy by means of the
definition $E'=E+k^2/2m$. Therefore, we recover the dispersion
relation $E'=p^2/2m$, which is physically consistent. Then we can
work with $k\neq 0.$

\section{Geometric Approach to Galilean Gravity}
 In order to generalize this formalism to a curved Galilean
space-time, we introduce Galilean tensor. Considering
\begin{equation}
\frac{\partial x^{\mu}\,'}{\partial x^{\nu}}=\Lambda^{\mu}\,_{\nu}
\,,\label{3.2}
\end{equation}
where $\Lambda^{\mu}\,_\nu $ is given explicitly by

\begin{equation}
\left(
  \begin{array}{c}
    \textbf{x}' \\
    x^4\,' \\
    x^5\,' \\
  \end{array}
\right)= \left(
           \begin{array}{ccc}
             R & 0 & -\textbf{V} \\
             -\textbf{V}\cdot R & 1 & \frac{1}{2}V^2 \\
             0 & 0 & 1 \\
           \end{array}
         \right)\,\left(
                    \begin{array}{c}
                      \textbf{x} \\
                      x^4 \\
                      x^5 \\
                    \end{array}
                  \right)
 \,, \label{3.3}
\end{equation}
we define covariant and contravariant components of tensors as
usual. Taking a non-flat manifold where locally the metric is
$\eta$,  we define a covariant derivative as
\begin{equation}
\nabla_\mu
X^{\nu}=\partial_{\mu}X^{\nu}+\Gamma^{\nu}_{\lambda\mu}X^{\lambda}
\,,\label{3.4}
\end{equation}
where $\Gamma^{\nu}_{\lambda\mu}$ is a connection that stipulates
the nature of Galilean space-time. The covariant derivative of a
scalar reduces to the normal derivative.

Let us give a  definition for the curvature tensor. The current
idea to define curvature resides on an intuitive concept. If we
consider a vector field $X^\mu$ on a closed circuit on a manifold
and if there is any change in direction of $X^\mu$ after a round
around the circuit then we say that this manifold is curved.
Mathematically it is stated as
\begin{equation}
\nabla_{[\mu}
\nabla_{\lambda]}X^{\nu}=\frac{1}{2}R^{\nu}\,_{\gamma\mu\lambda}X^{\gamma}
\,,\label{3.5}
\end{equation}
where $R^{\nu}\,_{\gamma\mu\lambda}$ is the curvature tensor. We
assume that when there is no gravitational field, the curvature
tensor vanishes.

The metric tensor is a covariant tensor of rank 2. It is used to
define distances and lengths of vectors. The infinitesimal
distance between two points  $x^a$ and $x^a+dx^a$ in curved
manifold defined from $\mathcal M$ is defined by
\begin{equation}
ds^2=g_{\mu\nu}dx^{\mu}dx^{\nu}\,,\label{3.6}
\end{equation}
where $g_{\mu\nu}$ is the metric tensor. The relation (\ref{3.6})
represents the line element as well. We have to notice that the
metric $\eta_{\mu\nu}$ in (\ref{1}) defines a flat line element.
The imposition that the covariant derivative of metric is zero
yields the following expression for the connection
\begin{equation}
\Gamma^{\nu}_{\lambda\mu}=\frac{1}{2}g^{\nu\delta}(\partial_{\lambda}
g_{\delta\mu}+\partial_{\mu}g_{\delta\lambda}-\partial_{\delta}g_{\lambda\mu})\,.\label{3.7}
\end{equation}
When the connection is written as in Eq.~(\ref{3.7}) the
manifold is said to be an affine manifold.

The curvature tensor defined in affine manifold has the following
properties:
\begin{eqnarray}
R_{\mu\nu\lambda\gamma}=-R_{\mu\nu\gamma\lambda}=-R_{\nu\mu\lambda\gamma}&=&
R_{\lambda\gamma\mu\nu}\nonumber \\
R_{\mu\nu\lambda\gamma}+R_{\mu\gamma\nu\lambda}+R_{\mu\lambda\gamma\nu}&\equiv&0.\label{3.8}
\end{eqnarray}
These properties are derived from Eq.~(\ref{3.5}). If we perform a
contraction of the indices of the curvature tensor then it is
possible to define the Galilei-invariant curvature scalar
\begin{equation}
R=g^{\mu\nu}g^{\gamma\lambda}R_{\gamma\mu\lambda\nu}.\label{3.9}
\end{equation}

To generate the field equations, we write a Lagrangian invariant
under Galilean transformations. A natural candidate is the
curvature scalar, then the action in a general form is
\begin{equation}
I=\int_\Omega d\Omega (\sqrt{-g}R+kL_m)\,,\label{3.10}
\end{equation}
where $g=det g_{\mu\nu}$, $k$ is the coupling constant, $L_m$ is a
matter lagrangian density and $d\Omega$ is the 5-dimensional
element of volume. Varying the action with respect to
$g_{\mu\nu}$, we obtain
\begin{eqnarray}
\frac{\delta(\sqrt{-g}R)}{\delta
g_{\mu\nu}}&=&-\sqrt{-g}(R^{\mu\nu}-\frac{1}{2}g^{\mu\nu}R)\,,\nonumber
\\
\frac{\delta L_m}{\delta
g_{\mu\nu}}&=&\sqrt{-g}T^{\mu\nu},\label{3.11}
\end{eqnarray}
where the latter equation defines the energy-momentum tensor of matter
fields and $R_{\mu\nu}=R^{\lambda}\,_{\mu\lambda\nu}$. Thus the
field equation becomes
\begin{equation}
R^{\mu\nu}-\frac{1}{2}g^{\mu\nu}R=kT^{\mu\nu}\,.\label{3.12}
\end{equation}
These equations have the same form as for the General Relativity
equations, since the Galilean transformations are written in a way
  similar to Lorentz transformations. Let us note that, the quantity $k$ is just
a coupling constant between the Galilean gravity and matter
fields  (it has nothing to do with Einstein's constant).

We have to note that the equations in (\ref{3.12}) with $\mu=4$,
i.e. the equations
\begin{equation}
R^{4}\,_{\nu}-\frac{1}{2}\delta^{4}_{\nu}R=kT^{4}\,_{\nu}\,,\label{3.13}
\end{equation}
contain only the first order derivative of the components of
$g_{\mu\nu}$ with respect to time. Actually  in (\ref{3.13}) the
components of the form $R_{4i4j}$ drop out, where the indices i and
j run from 1 to 3 and assume the value 5 as well. In face of this we note that some components of
time derivative of the metric tensor are associated with the freedom
of the choice of the system of coordinates. So we have to specify in
a particular coordinate system only the time derivatives of $g_{ij}$
as initial conditions. Therefore we see the ten equations
\begin{equation}
R^{i}\,_{j}-\frac{1}{2}\delta^{i}_{j}R=kT^{i}\,_{j}\,,\label{3.14}
\end{equation}
where the indices i and j run from 1 to 3 and assume the value 5 as well, as dynamical equations
and the five equations (\ref{3.13}) as constraint equations.

\section{The Schwarzschild-like Line Element:
The Advance of the Perihelion of a Planet}
\noindent

Since the metric determines every feature of a system described by a
geometrical approach, we intend to get a spherically symmetric
metric~\cite{Dinverno}. Then we introduce a Galilean Schwarzschild
solution. The Galilean Schwarzschild line element is
defined by
\begin{equation}
ds^2=f^{-1}(r)dr^2+r^2d\theta^2+r^2\sin^2\theta d\phi^2-f(r)dsdt\,,
\label{4}
\end{equation}
where as usual $f=1-2M/r$. This metric describes a system with
spherical symmetry. In the following we consider this  line
element  to study the movement of a planet.

The space-time in the exterior region of a massive body can be
described by the line element in Eq.~(\ref{4}). For example, for the
system composed of Sun and  Mercury, only force acting on this
system is the gravitational one.  Therefore, the movement will be
geodesic. A two-body system can be described by means of the reduced
mass as a consequence we deal with a one body system. We perform the
calculations considering the mass in the line element as the reduced
mass.

The geodesic movement can be described by means a variational
principle where the action is the interval between two events in
space-time. Then the equation of movement is given by
Euler-Lagrange equation, i.e.
\begin{eqnarray}
\delta s&=&\int\delta ds=\int\delta L d\tau\nonumber\\
&=&\int\delta
{(g_{\mu\nu}\dot{x}^{\mu}\dot{x}^{\nu})}^{1/2}d\tau=0\,,\label{4.01}
\end{eqnarray}
where the dot represents the derivative with respect to the proper
time $\tau$. The meaning of proper time remain the same of that defined in General Relativity, once our theory share the same feature with respect to transformation of coordinates. In this context the proper time is

\begin{equation}
\tau=\int{(-g_{00})}^{1/2}dt\,, \label{4.02}
\end{equation}
where the coordinate $t$ could not assume necessarily the meaning of
time. Thus the meaning of each coordinate depends on which system of
coordinates one is using.

Instead of working with $L={(g_{\mu\nu}\dot{x}^{\mu}\dot{x}^{\nu})}^{1/2}$, we consider the quantity defined by
\begin{equation}
K=g_{\mu\nu}\dot{x}^{\mu}\dot{x}^{\nu}\,,\label{4.1}
\end{equation}
which obey the Euler-Lagrange equation as well. Of course $K$ can be
set equal to a constant which takes the possible values 1, -1 or 0.
This relation is completely similar to that given in Eq.~(\ref{3.1})
defined on the flat Galilean space-time. If we perform the sum in
Eq.~(\ref{4.1}), the expression assumes the following form
\begin{equation}
f^{-1}\dot{r}^2+r^2\dot{\theta}^2+r^2\sin^2\theta\dot{\phi}^2-2f\dot{t}\dot{s}=\alpha
\,.\label{4.2}
\end{equation}

If we put $K$ into the Euler-Lagrange equation and consider this
movement restricted to a plane ($\theta=\pi/2$), since the angular
momentum is a constant, then we get for $\mu=(3,4,5)$ the
following equations, respectively,
\begin{equation}
r^2\dot{\phi}=h, \,\, f\dot{s}=\beta, \,\,
f\dot{t}=\kappa.\label{4.3}
\end{equation}
These equations represent conservation laws. In fact, the quantities
$\kappa$, $\beta$ and $h$ are related to energy, mass and angular
momentum respectively. The relations given in Eq.~(\ref{4.3}) can be
substituted into Eq.~(\ref{4.2}) with $\theta=\pi/2$, leading to
\begin{equation}
f^{-1}\dot{r}^2+\frac{h^2}{r^2}-2\frac{\beta\kappa}{f}=\alpha
\,.\label{4.4}
\end{equation}

At this point we  find an equation for the trajectory by changing
the variable in Eq.~(\ref{4.4}). We define
$U=U(\phi)=\frac{1}{r}$; such that the derivative of $U$ with
respect to $\phi$, which will be designated by $U'$, is equal to
$h$ times the derivative of $r$ with respect to the proper time.
The final equation is,
\begin{equation}
U^{'}\,^{2}+fU^2-2\frac{\kappa\beta}{h^2}=\frac{\alpha f}{h^2}
\,.\label{4.5}
\end{equation}
Taking the derivative of the above equation with respect to $\phi$
and remembering that $f=1-2M/r$, we obtain the trajectory equation,
\begin{equation}
U^{''}+ U=3MU^2-\frac{\alpha M}{h^2}.\label{4.6}
\end{equation}
Choosing $\alpha=-1$,  we have
\begin{equation}
U^{''}+ U= 3MU^2+\frac{M}{h^2} \,,\label{4.7}
\end{equation}
which is the same equation obtained in General Relativity. In this
context, however, there does not exist a relation of causality due
to the constancy of velocity of light, since  there is no such
imposition in a Galilean theory. As a consequence, the advance of
perihelion of a planet is given by the usual
expression~\cite{Landau}
\begin{equation}
\delta\phi=6\pi\frac{M^2}{h^2} \,. \label{4.8}
\end{equation}

It is important to rewrite Eq.~(\ref{4.8}) with the constants $G$
and $c'$. In this case we have
\begin{equation}
\delta\phi=6\pi\frac{G^2M^2}{c^{'2}h^2}=\frac{24\pi^3a^2}{c^{'2}T^2(1-e^2)}
\,, \label{4.9}
\end{equation}
where $c'$ is a typical velocity of the system to be fixed
experimentally  for the Galilean  symmetry, $M$ is the reduced mass, $T$ is the period of the movement,
$e$ is the eccentricity of the orbit and $a$ is semi-major axis of
the ellipse. We have to note that $\delta\phi$ is dimensionless.

In order to get the well known Newtonian equation for the
planetary movement under the influence of a force proportional to
the inverse of square radius, we have to rewrite Eq.~(\ref{4.7})
with the constants $G$ and $c'$, such that
\begin{equation}
U^{''}+ U=G\frac{M}{h^2}+ 3\frac{G}{c^{'2}}MU^2 .\label{4.10}
\end{equation}
This assumes the post-Newtonian equation for the planetary
movement, up to the second term on right-hand-side of above
equation. This term is a correction of the classical equation and
the parameter $c'$ can be taken experimentally. Therefore the
post-Newtonian equation arises from the 5-dimensional Galilean
gravity if we consider the parameter to adjust dimensionality
greater than the other parameters in Eq.~(\ref{4.10}); thus
establishing a weak field regime. It is important to notice that
Eq.~(\ref{4.9}) gives a precession of $43''/century$ for planet
Mercury,  since the parameter $c'$ is taken to be velocity of
light c. The consistence of this choice can be established by
noticing that, taking the Lorentz definition of energy and the
limit of $v/c$ much less than 1, then energy and momentum are
related by $E=p^2/(2m) +mc^2$, corresponding to a dispersion
relation as given in Section 2, $p_{\mu }p^{\mu }=p^{\mu }p^{\nu
}{\eta}_{\mu \nu }=\mathbf{p}^{2}-2p^{4}p^{5}=k^{2},$ with
$k^2=2(mc)^2$.

\section{Conclusion}
\noindent

We have established a geometric theory of gravity in the framework
of Galilean symmetry. These transformations are taken in a
covariant form and the flat space-time is defined with the
Galilean metric. With curved space-time we have associated this to
the gravitation, in parallel to the usual relativistic case. Then
we construct a Schwarzschild-like line element, which is
spherically symmetric, and apply it to the movement of a planet.
As a result, we derive the post-Newtonian result of the general
relativity in a covariant form.  We have shown that our equation
assumes the Newtonian form in the weak field regime. One basic
physical new fact to be learned from all this is that, the advance
of perihelion of a planet is a geometric effect that can be
described fully in a Galilean covariant theory.

It would be of interest to analyze other related problems in this
context of Galilean gravity, as for instance the bending of light,
other gravitation models such as cosmological models~\cite{Ehlers},
 the hamiltonian approach which could
reveal more about the structure of field equations and a Galilean gauge theory. These aspects
will be discussed elsewhere.

\bigskip
\noindent {\bf Acknowledgement}\par \noindent We thank R. Cuzinatto and P. Pomp\'{e}ia for helpful discussions. This work was
supported  by CNPq (of Brazil) and NSERC (of Canada).

%\bibliography{GGravityfev09}
%\bibliographystyle{apsrev}

\end{document}